                                                                  \documentclass[12pt,preprint]{aastex}
\shorttitle{The formation of  an inverse S-shaped active-region filament}
\shortauthors{Yan et al.}

\begin{document}

\title{The formation of an inverse S-shaped active-region filament driven by sunspot motion and magnetic reconnection}
\author{X. L. Yan\altaffilmark{1}, E.R. Priest\altaffilmark{2}, Q.L. Guo\altaffilmark{3}, Z. K. Xue\altaffilmark{1, 4}, J. C. Wang\altaffilmark{1}, L. H. Yang\altaffilmark{1, 4}}

\altaffiltext{1}{Yunnan Observatories, Chinese Academy of Sciences, Kunming 650011, China. yanxl@ynao.ac.cn}
\altaffiltext{2}{Mathematics Institute, University of St Andrews, St Andrews, KY16 9SS, UK.}
\altaffiltext{3}{College of Mathematics Physics and Information Engineering, Jiaxing University, Jiaxing 314001, China.}
\altaffiltext{4}{Key Laboratory of Solar Activity, National Astronomical Observatories, Chinese Academy of Sciences, Beijing 100012, China.}

\begin{abstract}
We present a detailed study of the formation of an inverse S-shaped  filament prior to its eruption in active region NOAA 11884 from October 31 to November 2, 2013. In the initial stage, clockwise rotation of a small positive sunspot  around the main negative trailing sunspot  formed a curved filament. Then the small sunspot  cancelled with negative magnetic flux to create a longer active-region filament with an inverse S-shape.  At the cancellation site a brightening was observed in UV and EUV images and bright material was transferred to the filament. Later the filament erupted after cancellation of two opposite polarities under the upper part of the filament. Nonlinear force-free field (NLFFF) extrapolation of vector photospheric fields suggests that the filament may have a twisted structure, but this cannot be confirmed from the current observations.

\end{abstract}


\section{Introduction}
Solar prominences (or filaments) are common features in the solar atmosphere and are one hundred times cooler and denser than the coronal material (Vr$\check{s}$nak et al. 1988; Tandberg-Hanssen 1995).  When they are viewed on the solar disk, they show strong absorption in H$\alpha$ and in the Extreme Ultraviolet (EUV) continuum, and are called `` filaments". When they are seen above the solar limb, they are bright features against a dark background, and are called ``prominences". Often, the terms ``filament" and``prominence" are used interchangeably. Prominence plasma is embedded in special magnetic structures and thermally isolated from the surrounding environment. It is formed in ``filament channels", which are the regions in the chromosphere surrounding a Polarity Inversion Line (PIL) where chromospheric fibrils are aligned with the PIL (Martin et al. 1994; Martin 1998a; Mackay et al. 1997, 2010; Gaizauskas et al. 1997, 2001; Kong et al. 2015). When they appear in active regions, at the border of active regions, and on the quiet Sun, they are divided into active-region filaments, intermediate filaments, and quiescent filaments, respectively (Parenti 2014). High resolution observations reveal that prominences consist of many thin threads with various material flows (Engvold 1998; Lin et al. 2005; Berger et al. 2008; Lin 2011; Yan et al. 2015a). In general, the height of quiescent filaments is larger than that of active-region filaments. But difference of physical nature between quiescent filaments and active-region filaments is not clear.

The fascination of prominences is their role as a crucially important component of solar activity. Generally, eruptions of prominences are associated with flares and coronal mass ejections (CMEs) (Shen et al. 2011; Deng et al. 2013; Kumar et al. 2013; McCauley et al. 2015).  Several observational precursors of filament eruptions have been found and several models have been proposed (see e.g. Priest 2014), but our focus here is on the formation process.

The formation of prominences (filaments) is still a challenging problem in solar physics. To address this issue, researchers have suggested models (based on observations and simulations), which are broadly divided into two types: one is surface effects that reconfigure coronal fields, the other is subsurface effects. Surface effects include shear flows along the PILs, differential rotation, converging flows towards PILs, and magnetic cancellation (van Ballegooijen \& Martens 1989; DeVore \& Antiochos 2000; Cheng et al. 2015). Subsurface effects involve the emergence of a twisted flux tube by magnetic buoyancy through the convective layer and its emergence into the photosphere, chromosphere, and corona, dragging cool dense material with it (Low 1994; Rust \& Kumar 1996). Active-region filaments are short-lived and low-lying compared with quiescent prominences. Observations of quiescent prominences tend to support surface flow mechanisms (Litvinenko \& Somov 1994; Chae et al. 2001; Wang \& Muglach 2007). Although an emergence mechanism has been proposed for active-region prominences (Lites \& Low 1997; Okamoto et al. 2008, 2009), we here present observational evidence for a combination of surface motions and magnetic cancellation being responsible.

Observations of the formation of solar filaments are relatively rare. Gaizauskas et al. (1997) have observed the early stages in the development of a filament channel and the subsequent formation of a filament in it. They found that the filament formed after the filament channel was created and filament formation was associated with a convergence between opposite magnetic polarities in a filament channel. Chae et al. (2001) have found that shear motion, converging motion, and magnetic cancellation   were closely related to the formation of a filament in active region NOAA 8688. Cancellation of small bipolar magnetic regions in a filament channel was also observed during the formation of solar filaments (Schmieder et al. 2004). For three case studies, Wang \& Muglach (2007) found that magnetic cancellation is the dominant process in the formation of filaments and their channels. Martin (1998b) has presented several observations relevant to the conditions for filament formation, such as their location at a boundary between opposite polarity magnetic fields, the convergence of opposite polarity network magnetic fields towards their common boundary within the channel, and cancellation of magnetic flux at the common polarity boundary. Recently, high-resolution observation from NVST have revealed that two successive active region filaments were formed by shear flow between opposite polarity magnetic fields and the rotation of a small sunspot in which one foot of the filaments was rooted (Yan et al. 2015b). A similar result was also presented by Yang et al. (2015). They found that a rotating magnetic field in the quiet Sun resulted in the formation of a circular filament. Yang et al. (2016) have also presented a case of filament formation, in which rapid formation of the filament was due to magnetic reconnection between two sets of dark threadlike structures. In addition, converging flows and continual flux cancellation have played important roles in filament formation.

In this paper,  we study the formation of an S-shaped active-region filament (in AR NOAA 11884) which erupts soon afterwards by using multi-wavelength observational data. Observations and methods are presented in Section 2. The results are shown in Section 3. The discussion are given in Section 4.

\section{Observations and methods}

Active-region NOAA 11884 produced several C-class and M-class flares during the period from October 31 to November 2, 2013. 
During its evolution, an inverse S-shaped filament was observed to form in this active region. Observations by the New Vacuum Solar Telescope (NVST; Liu et al. 2014) and the Solar Dynamics Observatory (SDO; Pesnell et al. 2012) covered the whole process of this formation. 

The Atmospheric Imaging Assembly (AIA; Lemen et al. 2012) on board the SDO provides multiple, simultaneous high-resolution full-disk images from the transition region to the corona. Full-disk UV and EUV images are taken by the AIA  with a 12-sec cadence and a spatial resolution of 0.$^\prime$$^\prime$6  per pixel (Lemen et al. 2012). The 304 \AA\ (HeII),171 \AA\ (Fe IX), 193 \AA\ (Fe XII, XXIV) images observed by SDO/AIA were used to show the  formation of the active-region filament. We also used full-disk line-of-sight magnetograms at 45 s cadence with a precision of 10 G and vector magnetic fields in the photosphere observed by the Helioseismic and Magnetic Imager (HMI; Schou et al. 2012; Bobra et al. 2014; Centeno et al. 2014). The vector magnetograms came from Space Weather HMI Active Region Patch (SHARP) series, which have a pixel scale of about 0.$^\prime$$^\prime$5 and a cadence of 12 minutes. They were derived using the Very Fast Inversion of the Stokes Vector algorithm (Borrero et al. 2011). The minimum energy method (Metcalf 1994; Leka et al. 2009; Metcalf et al. 2006) was used to resolve the 180 degree azimuthal ambiguity. The images were remapped using Lambert (cylindrical equal area) projection centered on the midpoint of the AR, which is tracked at the Carrington rotation rate (Sun 2013). 

The NVST is a vacuum solar telescope with a 985 mm clear aperture and is located in Fuxian Lake of China. It is a multi-channel high-resolution imaging system (one channel for the chromosphere (6563 \AA) and two channels for the photosphere (TiO 7058 \AA\ and G-band 4300 \AA) and spectrometer systems (a multi-band spectrometer and a high-dispersion spectrometer). The H$\alpha$ images used in this paper from the NVST were taken at H$\alpha$ center (6563 \AA) with a bandwidth of 0.25 \AA . They have a pixel size of 0.$^\prime$$^\prime$163 and a cadence of 12 s. The data are calibrated from Level 0 to Level 1 with dark current subtracted and flat field corrected, and then the calibrated images are reconstructed to Level 1+ by speckle masking (Weigelt 1977; Lohmann et al. 1983).  We used a cross-correlation method to co-align the H$\alpha$ images. The method is divided into three steps: one is to remove the motion of the full-field images from one frame to the next by using a cross-correlation method; the next is to divide the full field images into a number of sub-images and the sub-images were co-aligned using the cross-correlation method. Moreover, the sub-images were reconstructed from subsequent images. Finally, all the reconstructed sub-images are combined to form the whole high resolution image (see Liu et al. 2014; Xiang et al. 2016).

Moreover, we used vector magnetic fields from SDO/HMI as boundary conditions to extrapolate the magnetic fields from photosphere to corona. The optimization algorithm proposed by Wheatland et al. (2000) and implemented by Wiegelmann (2004) was employed to extrapolate the three-dimensional NLFFF structure. We applied a preprocessing procedure (fff\_temp\_pre.pro in SSW) to the bottom boundary vector data before extrapolation. This procedure ensures the observational vector magetograms fulfill the following conditions: (i) The data coincides with the photospheric observations within measurement errors; (ii) The data is consistent with the assumption of a force-free magnetic field; (iii). the data is smooth. This removes most of the net force and torque that produce an inconsistency between the observed photospheric magnetic field and the force-free assumption (see Wiegelmann et al. 2006 for details).

The AIA and HMI data were calibrated to Level 1.5 by using the standard procedure in SSW, and rotated differentially to a reference time (at 04:30:07UT on November 1, 2013). Then, we co-aligned the SDO and NVST images using the subpixel registration algorithm, respectively (Feng et al. 2012; Yang et al. 2014; Ji et al. 2015). Because 304 \AA\ and H$\alpha$ images have similar structures, we used a cross-correlation method to align the two images and then the other wavelength images of AIA were co-aligned with the 304 \AA\ images. The error of co-alignment is estimated to be about 0.$^\prime$$^\prime$6.

\section{Results}
\subsection{The formation of the inverse active-region filament}
During the period from October 31 to November 2, 2013, active region NOAA 10884 was located near the center of the Sun with a sunspot group having a $\beta$$\gamma$$\delta$ magnetic configuration. Figure 1 shows the formation of the active-region filament in line-of-sight magnetograms, 304 \AA\ images, and H$\alpha$ images with the line-of-sight magnetograms superimposed. Blue and red contours indicate negative and positive magnetic fields with contour levels $\pm$500 G and $\pm$1000 G. Green arrows in Figs. 1(a1)-(a3) indicate the sunspot (C) with positive polarity, which was rotating around a trailing sunspot (D) with negative polarity. The rotation angle was about 40 degrees during 33 hours. This rotating sunspot was of type II (Yan et al. 2008). The chromospheric fibrils connecting the rotating sunspot were dragged to form a curved filament (CD) marked by white arrows in Figs. 2(b1)-(b3) and Figs. 1(b1)-(b3), and outlined by yellow dotted lines marked 1 in Figs. 1(c1)-(c3). The lower foot of the curved filament is rooted in the rotating sunspot (C). Meanwhile, the footpoint (B) of chromospheric fibrils (AB) (marked 2 in Figs. 1(c1)-(c3)) were moving from north-east to south-west. Eventually, the filament footpoint B combined through magnetic cancellation with the arcade footpoint C and formed a long S-shaped filament AD (see Figs. 1(b4) and (c4)). The formation process can be seen in online animated Figure 1. The red and blue contours outline the positive and negative polarities. The contour levels are  $\pm$500 G, $\pm$1000 G, and $\pm$1500 G. Note that the dark H$\alpha$ fibrils that were moving from east to west in the field of view (FOV) during the period from 01:22UT to 04:10UT on Nov. 01, 2013 are not related to the filament formation studied in this paper. The H$\alpha$ images and magnetograms are linearly scaled while 304 \AA\ images are on a log scale. From the movie, the motion of the sunspot with positive polarity (indicated by the green arrows in Figs. 1(a1)-(a3)) appears to have dragged the chromospheric fibrils and given them a curved structure. Meanwhile, the brightening at cancellation site can also be seen in H$\alpha$ observation. The formation of the inverse active-region filament was observed in several wavelengths of SDO/AIA. Figure 2 presents the observations acquired in 304 \AA\, 171 \AA\, and 193 \AA. The white arrows indicate the curved filament (a1-a3, b1-b3, and c1-c3) and the inverse S-shaped filament (a4, b4, and c4). The regions marked by the blue boxes are the same as those marked by the blue box in Fig. 1(b3). The formation process is very similar to the observation acquired at 304 \AA\ wavelength. The online animated Figure 2 shows the formation of the S-shaped filament in 304 \AA, 171\AA, and 193 \AA\ from 07:59:59 UT on 2013 October 31 to 07:35 UT on November 3, including two eruptions of the filament that occurred shortly after its formation (which are not discussed in this paper).

\subsection{Magnetic cancellation}
In order to show more clearly the change of the magnetic fields, their evolution in the green box in Fig. 1(a3) is shown in Fig. 3 from 08:00:00UT on October 31 to 23:36:00 UT on November 2, 2013. The two blue boxes marked 1 and 2 are used to show the cancellation of two strings of opposite polarity, the moving magnetic flux with negative polarity and the sunspot with positive polarity, respectively. The yellow lines indicate positive polarities, while the green lines indicate negative polarities. The blue boxes 1 and 2 are the same as that in Fig. 1(a3). During the formation of the filament, the two opposite polarities approached each other. The string with negative polarity exhibited an arcade shape around the trailing sunspot. Ultimately, they cancelled and the positive polarities disappeared completely. Fig. 4(a) shows the movement of two opposite polarities along the red line marked in Fig. 3(a). The positive polarity moved a little with a speed of 16 m/s. The velocity of the negative polarity approaching the positive polarity was 126 m/s. The change of magnetic flux in box 1 is shown in Fig. 4(b). The red and blue lines indicate the change of positive and negative magnetic flux, respectively. The two vertical dashed lines show the duration of cancellation, which began at about 16:00 UT on November 1, 2013. The positive polarity was completely cancelled with the negative polarity by 20:00 UT on November 2, 2013. The region in box 2 also experienced cancellation (see Figs. 3(b)-3(f)).  From 13:12:00 UT on November 1 to 00:12:00 UT on November 2, 2013, the negative polarity cancelled with the sunspot with positive polarity. The change of magnetic flux in box 2 is shown in Fig. 4(c). The red and blue lines indicate the change of positive and negative magnetic flux, respectively. During the magnetic cancellation (the period marked by two vertical dashed lines in Fig. 4(c)), the cancellation site exhibited brightening in the chromosphere. Fig. 4(d) shows the change of 304 \AA\ normalized intensity in the blue box 2 in  Fig. 1(b3). There are several discontinuous changes in emission in 304 \AA\ images during magnetic cancellation (see the period between the two vertical dashed lines in Fig. 4(d)). The cancellation process can be seen in the online animated Figure 3. The two blue boxes mark the two regions that exhibited significant magnetic cancellation. The white and black patches indicate positive and negative polarity magnetic fields. Magnetic fields between - 800 G and 800 G are presented in Fig. 3.

\subsection{NLFFF extrapolation}
The vector magnetograms and NLFFF extrapolation are shown in Fig. 5. The left panel shows the evolution of the vector magnetic fields from October 31 to November 1, 2013, while the right panel shows the resulting NLFFF extrapolation. The blue arrows represent transverse magnetic fields in Figs. 5(a1)-(a3). The structure of the filament during its formation can be seen in Figs. 5(b1)-(b3). The filament gradually formed a curved structure (CD) (see Figs. 5(b1)-(b2)). Then it reconnected with the arcade fibrils (AB). Finally, an inverse S-shaped filament (AD) was formed (see Fig5. b3). The curved filament (CD) and the inverse S-shaped filament (AD) both exhibited a twisted flux rope structure. 

The presence of twist in a flux rope before eruption is highly likely on theoretical grounds, since an increase in twist adds magnetic energy in excess of potential to a flux tube and makes it more likely to become unstable and erupt (e.g., Priest, 2014). Furthermore, in the present example, there are several ways of building up the twist. First of all, vortex motions associated with the rotation of the small positive spot around the main negative one will tend to add twist to the flux rope.  Secondly, the two cancellation locations are likely to add twist by three-dimensional reconnection, during which the total magnetic helicity is conserved, so that mutual magnetic helicity can be converted into self magnetic helicity. Wang and Muglach (2007) considered the formation of filaments by flux cancellation acting on a succession of pairs of loops, and suggested that this could produce a sheared field without twist. The newly formed loops in the model proposed by Wang and Muglach (2007) have increased shear, whereas the model proposed by van Ballegooijen \& Martens (1989) suggests that the reconnected loops are twisted. Both of these cartoon models are possible formation mechanisms using flux cancellation, but detailed modelling is needed to determine which is more appropriate in the present case.

Observationally, twist is often seen in erupting prominences, but is notoriously difficult to detect in filaments before eruption and even more difficult to quantify, partly because of the lack of resolution of dark filamentary structure.  In our case, there are suggestions of twist in the filament in the early stages in Figure 1 (b1, b2, b3, c3) and Figure 2 (a2, a3) and also in the later stages after cancellation in Figures 6 (b4) and Figure 7 (a4 and b4). But we cannot be certain from the observations that the pre-eruptive filament is twisted.

\subsection{Plasma injection}
During the magnetic cancellation in the region marked by the blue box 2 in Figs.1(a3), (b3), Fig. 2, and Fig. 6, the brightening first appeared at the cancellation site and then the heated plasma was injected into the filament channel. Four examples are shown in Fig. 6. The upper panels show the onset of brightening at the cancellation site, while the lower panels show the corresponding change of the filament structure after cancellation. Bright plasma was injected into the filament from the cancellation site and the whole filament was heated by it (see Figs. 6(b1)-(b4)). Brightening of the active-region filament in 171 \AA\ and 193 \AA\ images was also observed after the magnetic cancellation during 04:19 UT on Nov. 1 to 08:08 UT on Nov. 2 (see Fig. 7). Plasma injection in filament channels due to flux cancellation has recently been observed and discussed by Wang \& Muglach (2013). They suggested that elongated brightenings result from reconnection between pairs of horizontal loops that extend outward in opposite directions along the filament channel and have one footpoint located in the region of canceling flux. In this event, the injection of the hot plasma was due to magnetic reconnection between a curved filament and the chromospheric fibrils.

\subsection{Lengthening the filament by reconnection}
A simple model to show how reconnection can lengthen a filament is as follows. Suppose a flux tube of flux F stretches from a point A$(0,0)$ to a point  B$(2,0)$ on the x-axis, and that another flux tube also of flux F stretches between C$(c,0)$ and D$(4,0)$, so that we are measuring lengths in terms of twice the distance between A and B. Reconnection between the two tubes to create a longer tube joining  A and D may be modelled by considering evolution through a series of two-dimensional potential fields as the parameter c decreases from 4 to 2 (Fig. 8).

The magnetic field components may be written
\begin{eqnarray}
{B_r}=\frac{\partial A}{r \partial \theta},\ \ \ \ \ \ \ B_\theta=-\frac{\partial A}{\partial r}
\label{eq1}\
%\nonumber
\end{eqnarray}
in terms of the flux function ($A$). For a point source at the origin that produces a flux F in the upper half-plane, the magnetic field is 
$B_r=F/(\pi r)$, while the flux function is $A=F\theta/\pi$.

Suppose there are sources of fluxes $+F$, $-F$, $+F$ and $-F$ at A, B, C, D, respectively. Then, using the notation in Fig. 8a, the magnetic field components are
\begin{eqnarray}
B_x=\frac{F}{\pi}\left (\frac{\cos \theta}{r}-\frac{\cos \theta_B}{r_B}+\frac{\cos \theta_C}{r_C}-\frac{\cos \theta_D}{r_D}\right),
\label{eq2}\\
B_y=\frac{F}{\pi}\left(\frac{\sin \theta}{r}-\frac{\sin \theta_B}{r_B}+\frac{\sin \theta_C}{r_C}-\frac{\sin \theta_D}{r_D}\right),
\label{eq3}
\end{eqnarray}
where $\cos \theta=x/r$, $\cos \theta_B=(x-2)/r_B$, $\cos \theta_C=(x-c)/r_C$, $\cos \theta_D=(x-4)/r_D$,  
$\sin \theta=y/r$, $\sin \theta_B=y/r_B$, $\sin \theta_C=y/r_C$, $\sin \theta_D=y/r_D$,  
$r^2=x^2+y^2$, $r_B^2=(x-2)^2+y^2$, $r_C^2=(x-c)^2+y^2$, $r_D^2=(x-4)^2+y^2$.

There is a null point (N) where the magnetic field vanishes at ($x_N,y_N$), say, given by setting $B_x=B_y=0$ in Eqs. (2) and (3), namely,
\begin{eqnarray}
\frac{x}{x^2+y^2}-\frac{x-2}{(x-2)^2+y^2}+\frac{x-c}{(x-c)^2+y^2}-\frac{x-4}{(x-4)^2+y^2}=0,
\label{eq4}\\
\frac{1}{x^2+y^2}-\frac{1}{(x-2)^2+y^2}+\frac{1}{(x-c)^2+y^2}-\frac{1}{(x-4)^2+y^2}=0.
\label{eq5}
\end{eqnarray}
The way in which the coordinates of N change as $c$ decreases from $c=4$ to $c=2$ and C moves from D to B is shown in Fig. 8b. Combining Eqs. (4) and (5), we find $x_N$ and $y_N$ to be
\begin{eqnarray}
x_N=\frac{8}{6-c},\ \ \ \ \ \ y_N={\frac{2\sqrt{2}}{6-c}}\sqrt{|c^2-6c+8|},
\label{eq6}
%\nonumber
\end{eqnarray}

The value of the magnetic flux going from C to B is
\begin{eqnarray}
F_{CB}=\int_0^{y_N} | B_x |_{x=x_N} dy
\label{eq7}
%\nonumber
\end{eqnarray}
Then, since the total magnetic flux entering B is $F$,  the magnetic flux that joins A and B is $F_{AB}=F-F_{CB}$. Also, since the flux leaving A is $F$, the flux joining A and D is $F_{AD}=F_{CB}$ (Fig.  8c). The way in which $F_{AB}$ decreases from $F$ to zero, while $F_{AD}$ increases from 0 to $F$ as $c$ decreases from 4 to 2 is shown in Fig. 9c. The curves of the relationship between $x_N$, $y_N$, and c can be seen in Figs. 9a, b.\\

\subsection{Changing the twist of a filament by reconnection}
The way in which reconnection of a filament can change its twist may be illustrated as follows.   If two flux tubes are in the same line, their twists just add together when they reconnect.  However, when they are inclined at some angle, then the net twist after reconnection may be different, due to the conversion of mutual helicity into self-helicity. Our example may be modelled by comparing the configurations just before and just after reconnection. 

Suppose the long flux tube (AB) of twist $\Phi_{AB}$ and flux $F$ has length 2 and is inclined at $\pi/4$ to the other tube (CD) of $\Phi_{CD}$, flux $F$ and length 1, as indicated in Fig. 10 and suppose C  is just about to coincide with B and create one long tube (AD) of twist $\Phi_{AD}$.  The initial self-helicities of AB and CD are
$\Phi_{AB}F^2/(2\pi)$ and $\Phi_{CD}F^2/(2\pi)$, respectively, and their mutual helicity is $(\theta_2-\theta_1)F^2/(2\pi)$. After reconnection a single tube AD is created with self-helicity $\Phi_{AD}F^2/(2\pi)$, so that conservation of helicity during reconnection implies that the sum of the self-helicities of AB and CD and their mutual helicity must equal the final self-helicity of AD, so that
\begin{eqnarray}
\Phi_{AD}=\Phi_{AB}+\Phi_{CD}+\theta_2-\theta_1,
\label{eq8}%\nonumber
\end{eqnarray}
where $\theta_2=\pi/4$ and $\theta_1=\tan^{-1}(\sqrt 2/(1+\sqrt 2))\approx(2\pi)/25$.
The measured twist values of two parts in the nonlinear force-free extrapolation are 0.38 and 0.6 (Fig. 5(b2)).  Therefore, $\Phi_{AD}$ equals 1.15. The final twist of the filament calculated from the nonlinear force-free extrapolation in Fig.5(b3) is 1.49. The twist value of the model is consistent with that of the extrapolation.

One important effect of lengthening a flux rope is to increase the height of its summit so much that it may become unstable to torus instability, which occurs when the summit height exceeds a critical value. Another effect is to increase its twist, so that the rope may become unstable to kink instability.

\section{Discussion}

During the evolution of active region NOAA 10484, it was found that a sunspot with positive polarity rotated around a trailing sunspot with negative polarity. Chromospheric fibrils connecting the sunspot with positive polarity were dragged to form a curved filament. After cancellation of the positive sunspot with the moving magnetic flux of negative polarity, the curved filament and arched fibrils combined together through magnetic cancellation and formed an inverse S-shaped filament. During the formation of the sigmoid filament, flux with opposite polarity moved together and cancelled under the upper part of the filament. According to NLFFF extrapolation, the formation of the sigmoid filament can be obtained from the evolution of the magnetic structures. Moreover, twisted magnetic structure for the filament is present in the extrapolation. In addition, hot plasma was found to be injected into the filament from the cancellation site. 

Two means for forming solar filaments have been suggested by previous researchers. One is the building of magnetic structures through photospheric flux cancellation and/or coronal magnetic reconnection (van Ballegooijen \& Martens 1989; DeVore \& Antiochoes 2000; Mackay et al. 1998; Yang et al. 2015). The other is the emergence of a flux rope from below the photosphere (Low 1994; Fan \& Gibson 2006; Rust \& Kumer 1994; Magara 2006; Xu et al. 2012). The observations of Martin (1998b) and Wang \& Muglach (2007) suggested that shearing motion, convergence and cancellation of magnetic flux play a critical role in the formation of filaments. The emergence of a horizontal flux rope from below the photosphere was also observed by Lites \& Low (1997) and Okamoto (2009). Moreover, more and more observations provide evidence that active-region filaments have twisted magnetic structures (Wang et al. 1996; Yan et al. 2013, 2014; Yang et al. 2014; Bi et al. 2015; Hou et al. 2016) as well as the extrapolations (Guo et al. 2010; Cheng et al. 2013; Xia et al. 2014; Wang et al. 2015). However, there is still controversy on the structures of filaments (Ouyang et al. 2015). Our observation is a little different from previous ones. Firstly, the formation of the curved filament was closely related to the rotation of a sunspot, which stretched out and twisted the field lines. The filament later evolved into a longer inverse S-shaped filament through magnetic reconnection with arcade fibrils. Reconnection was evidenced by magnetic cancellation in the photosphere. The simple cartoon in Fig. 11 shows the formation of the inverse S-shaped active-region filament. Minus and plus symbols indicate negative and positive polarities.

Formation of the curved filament was due to the sunspot rotation around the trailing sunspot. The convergence and cancellation of two opposite polarities started under the curved filament. The filament also exhibited movement toward the west. The negative polarity followed the movement of the filament. The cancellation of two opposite polarities continued after the formation of the inverse S-shaped filament. When the positive flux was cancelled completely, the filament began to erupt. The filament formation studied in this paper is consistent with a surface-effect mechanism, in which sunspot motion and magnetic cancellation play important roles. From the observations, we cannot deduce that the filament has a twisted magnetic structure. The later eruption needs to be studied in more detail to deduce whether it is due to torus or kink instability which require a twisted magnetic structure or due to mechanisms that do not need a twisted structure like breakout or tether cutting. However, the mechanism of flux emergence can be excluded in this event, since the data clearly shows flux cancellation before the filament eruption. 

There is a debate about the magnetic topology of solar filaments. One possibility is a twisted flux rope (Priest et al. 1989; Demoulin \& Priest 1989; Amari et al. 2000; Gibson \& Fan 2006; Jiang et al. 2014; Su et al. 2015; Li et al. 2015) and the other is a sheared arcade structure (Amari et al. 1991; Antiochos et al. 1994; DeVore \& Antiochos 2000; Welsch et al. 2005). For this event, the nonlinear force free extrapolation suggests that the filament has a twisted magnetic structure, but such a twisted structure cannot be seen clearly in the observations at present.

\acknowledgments 
The authors thank the referee for her/his constructive suggestions and comments that helped to improve this paper. SDO is a mission of NASA's Living With a Star Program. The authors are indebted to the SDO and NVST teams for providing the data. The authors thank Yongyuan Xiang for high-resolution reconstruction of H-alpha images.This work is supported by the National Science Foundation of China (NSFC) under grant numbers 11373066, 11503080, 11633008, Yunnan Science Foundation of China under number 2013FB086, CAS ``Light of West China" Program, Key Laboratory of Solar Activity of CAS under number KLSA201303, Youth Innovation Promotion Association CAS (No. 2011056).

\begin{figure}
\epsscale{.60}
\plotone{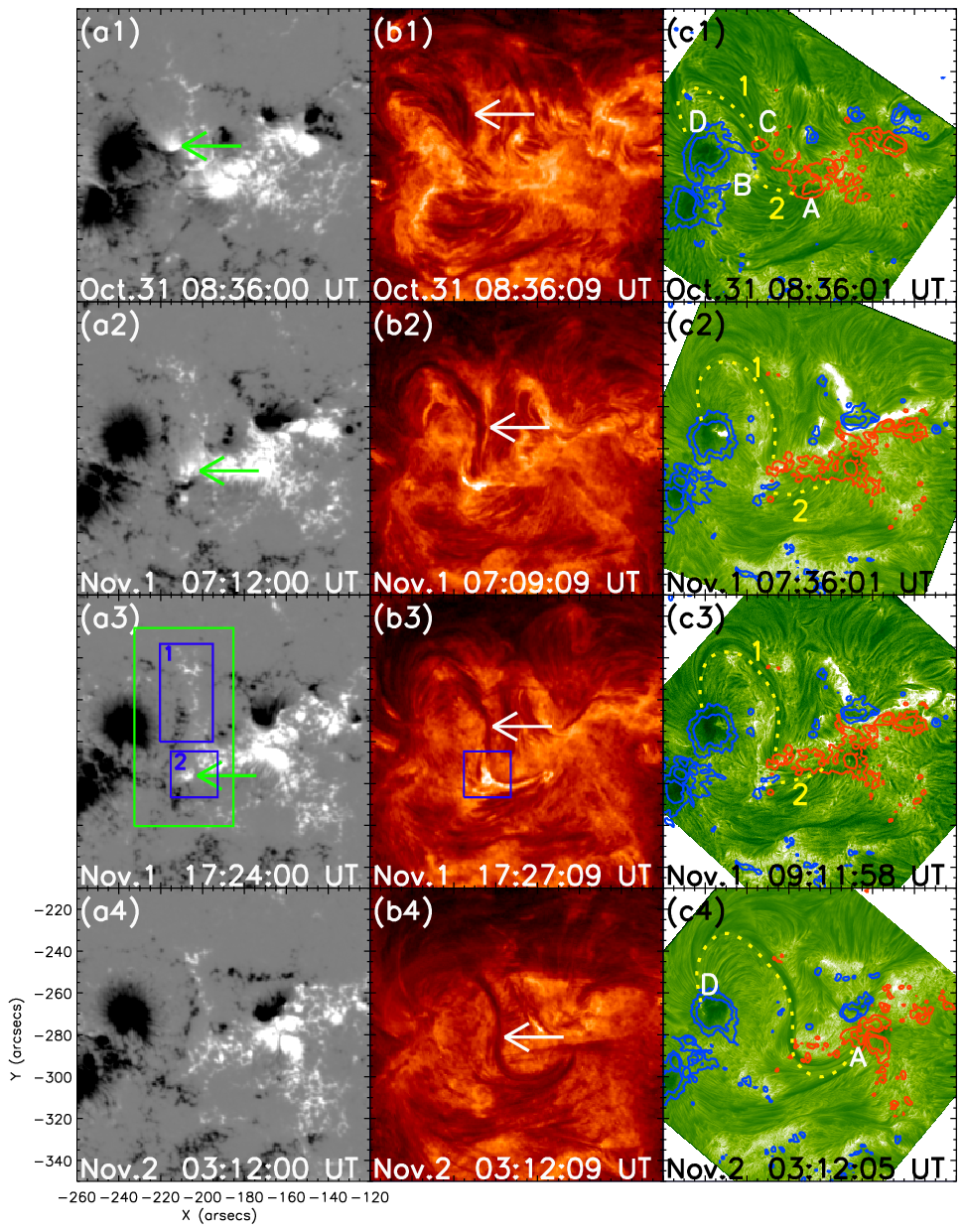}
\caption{The formation of an active-region filament observed by SDO and NVST from October 31 to November 2, 2013. Figs. 1(a1)-(a4) and Figs. 1(b1)-(b4) show the evolution of the line-of-sight magnetic fields in the photosphere observed by HMI and the structure of the active-region filament in the chromosphere observed by AIA at 304 \AA. Figs. 1(c1)-(c3) show the H$\alpha$ images with the line-of-sight magnetograms superimposed. Blue and red contours indicate positive and negative magnetic fields with contour levels $\pm$500 G and $\pm$1000 G. Green arrows indicate the sunspot (C) with positive polarity, while the white arrows indicate the active-region filament. The numbers 1 and 2 mark the curved filament (CD) and the arcade fibrils (AB). The green box in Fig. 1(a3) denotes the field of view of Fig. 2. Note that the dark H$\alpha$ fibrils from east to west in the FOV during the period from 01:22UT to 04:10UT on Nov. 01 are not related to the filament formation studied in this paper. An online animated version of panels (a) and (c) in this Figure shows the formation process of the filament. \label{fig1}}
\end{figure}

\begin{figure}
\epsscale{.50}
\plotone{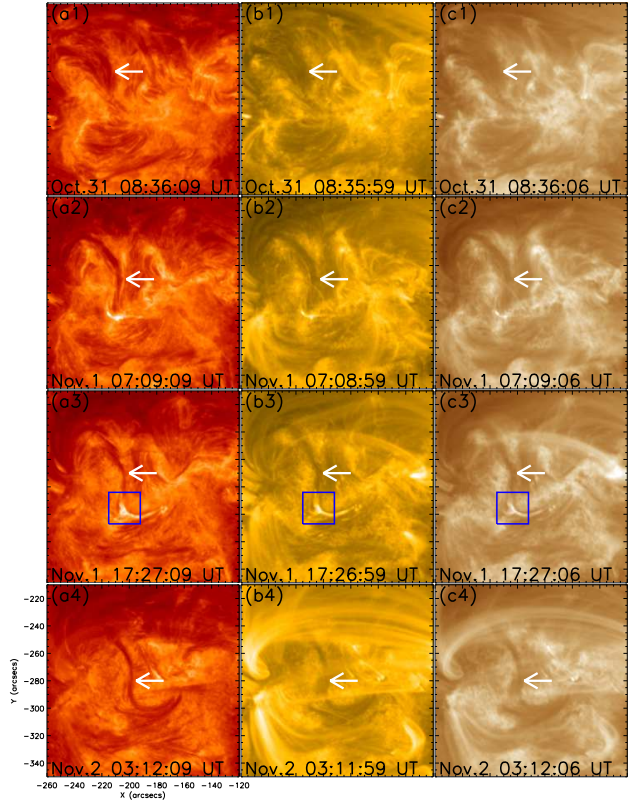}
\caption{The formation of the active-region filament acquired at 304 \AA\, 171 \AA\ and 193 \AA\ wavelengths. The white arrows indicate the curved filament (a1-a3, b1-b3, and c1-c3) and the inverse S-shaped filament (a4, b4,  and c4). The regions marked by the blue boxes are the same as those marked by blue boxes in Fig. 1(b3). An online animated version of this Figure shows the formation process of the filament. \label{fig1}}
\end{figure}

\begin{figure}
\epsscale{.80}
\plotone{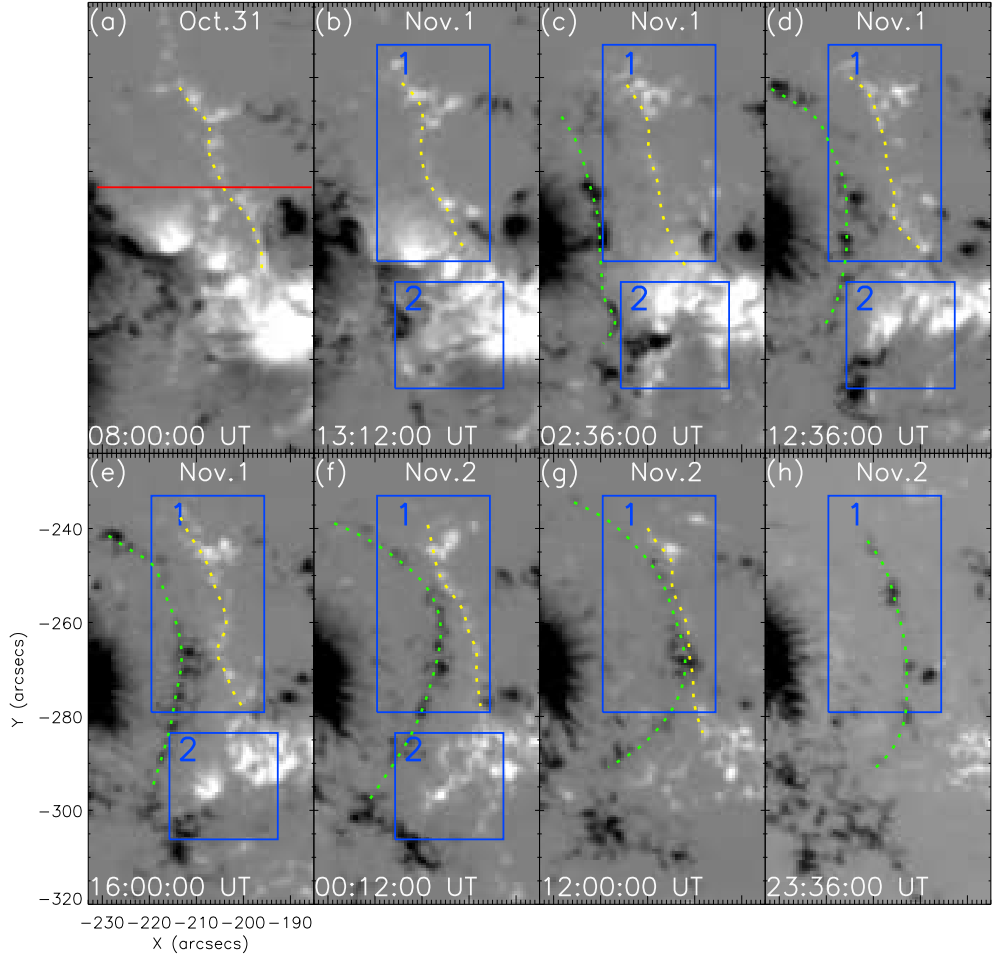}
\caption{The evolution of the magnetic fields during the formation of the active-region filament. 	Yellow and green lines indicate the strings of positive and negative polarities, respectively. The regions marked by the blue boxes 1 and 2 are the same as those marked by blue boxes in Fig. 1(a3). The field of view of Fig. 2 is marked by the green box in Fig. 1(a3). The red line in Fig. 3(a) shows the position of the time slice of Fig. 4(a). An online animated version of this Figure shows magnetic cancellation during the filament formation. \label{fig1}}
\end{figure}

\begin{figure}
\centering
\includegraphics[angle=0,scale=1.2]{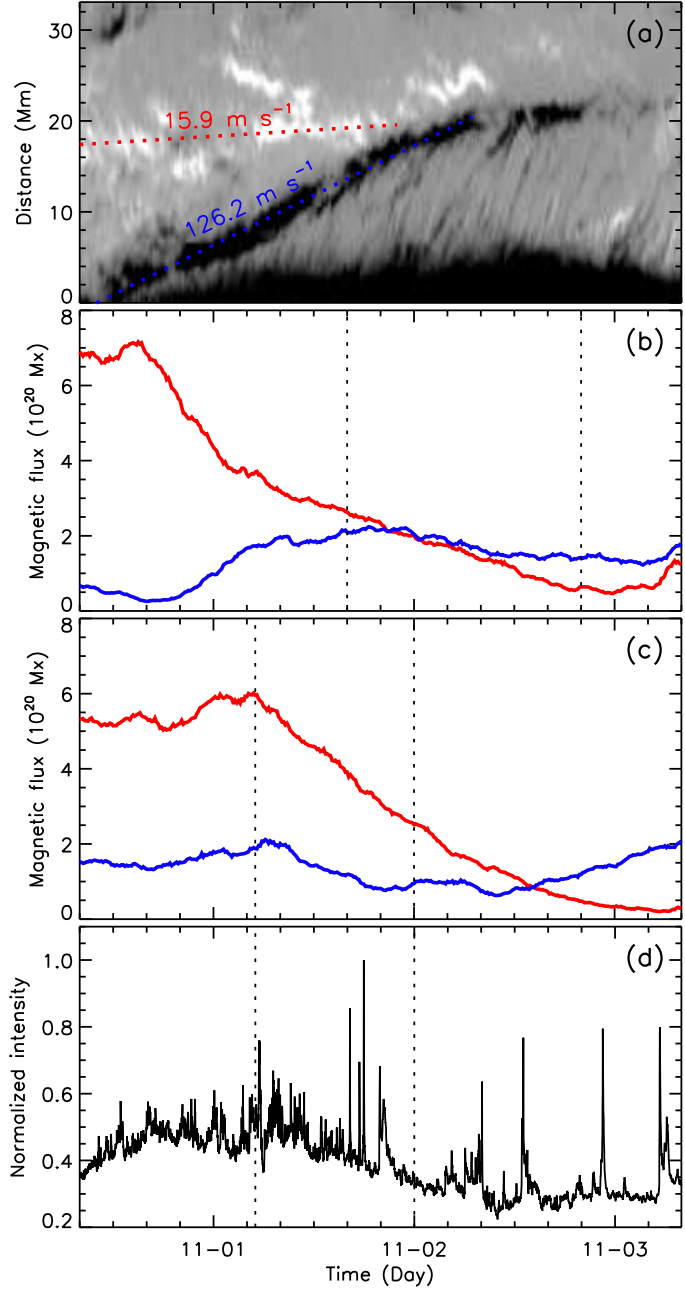}\\
\caption{Panel (a): A time slice along the red line in Fig. 3(a). Red and blue dotted lines show the movement of the positive and negative polarities along the red line in Fig. 2(a). Panel (b) and (c): Temporal evolution of magnetic flux in the regions marked by blue boxes 1 and 2, respectively. The red and blue lines indicate the evolution of the positive and negative magnetic flux. Panel (d): Temporal evolution of the 304 \AA\ normalized intensity in the blue box marked in Fig. 1(b3). The time period of these panels is from 08:00 UT on October 31 to 08:00 UT on November 3, 2013. The two vertical dashed lines (panels b, c, and d) show the duration of cancellation in box 1 and box 2, respectively.}
\end{figure}%\begin{figure}

\begin{figure}
\epsscale{.80}
\plotone{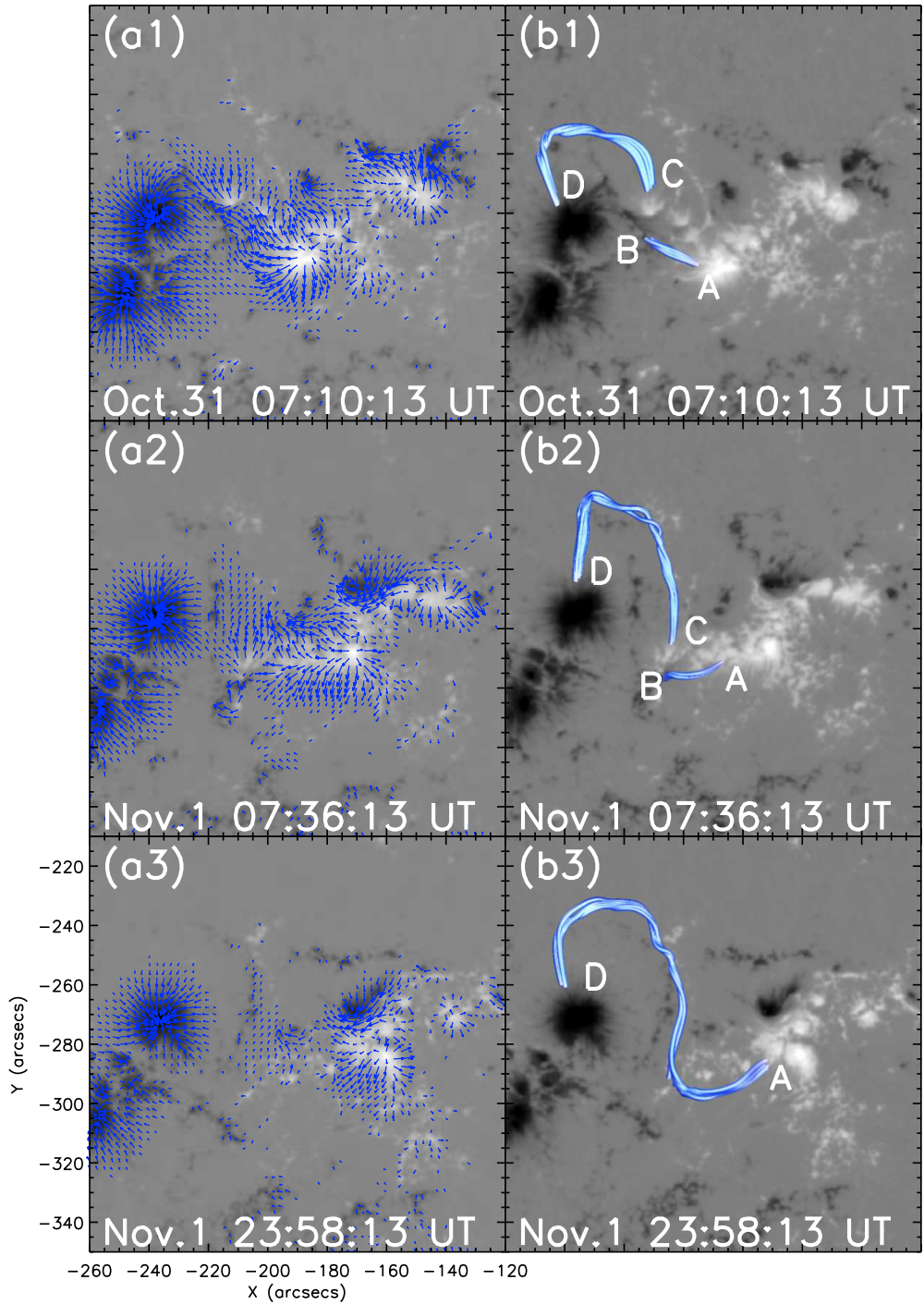}
\caption{The vector magnetic fields observed by SDO/HMI and 3D NLFFF extrapolations from October 31 to November 1, 2013. Left panel: The vector magetograms at different time. Blue arrows show the transverse magnetic fields. Right panel: Extrapolations of the filament structure superimposed on the longitudinal magnetic fields. \label{fig1}}
\end{figure}%

\begin{figure}
\epsscale{.80}
\plotone{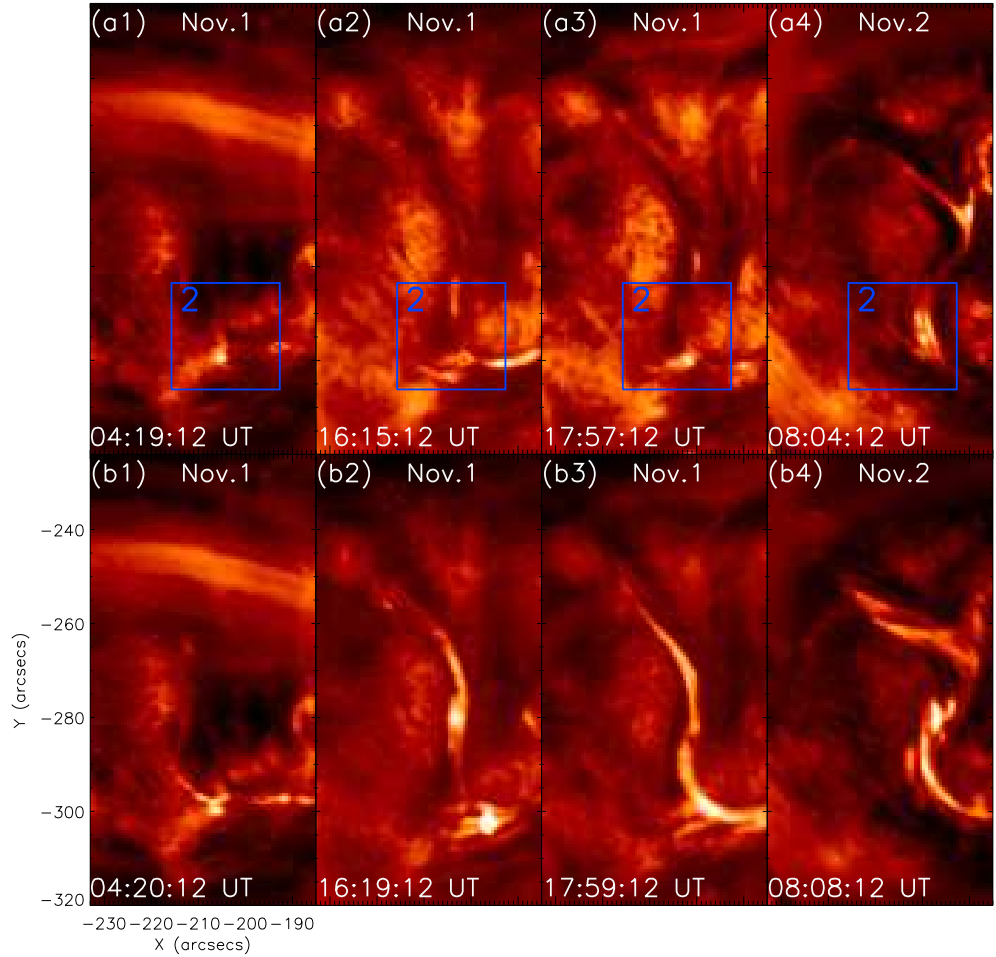}
\caption{Four views showing the change of structure of the active-region filament in 304 \AA\ images after brightening at the cancellation site. The upper panels show the onset of brightening at the cancellation site, while the lower panels show the corresponding change of the filament structure after cancellation. Box 2 is the same as that in Fig. 1(b3). \label{fig1}}
\end{figure}

\begin{figure}
\epsscale{.80}
\plotone{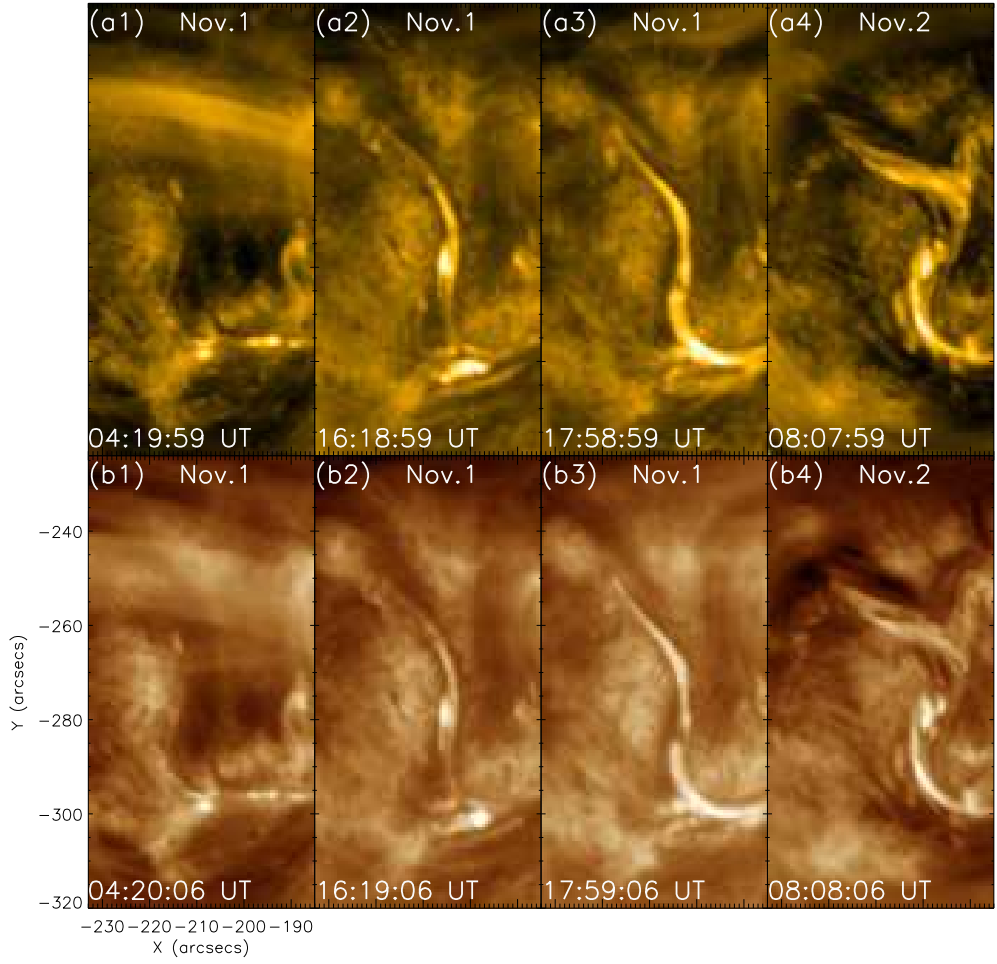}
\caption{Brightening of the active-region filament in 171 \AA\ and 193 \AA\ images after the magnetic cancellation. \label{fig1}}
\end{figure}

\begin{figure}[ht]
{\centering
 \includegraphics[width=10cm]{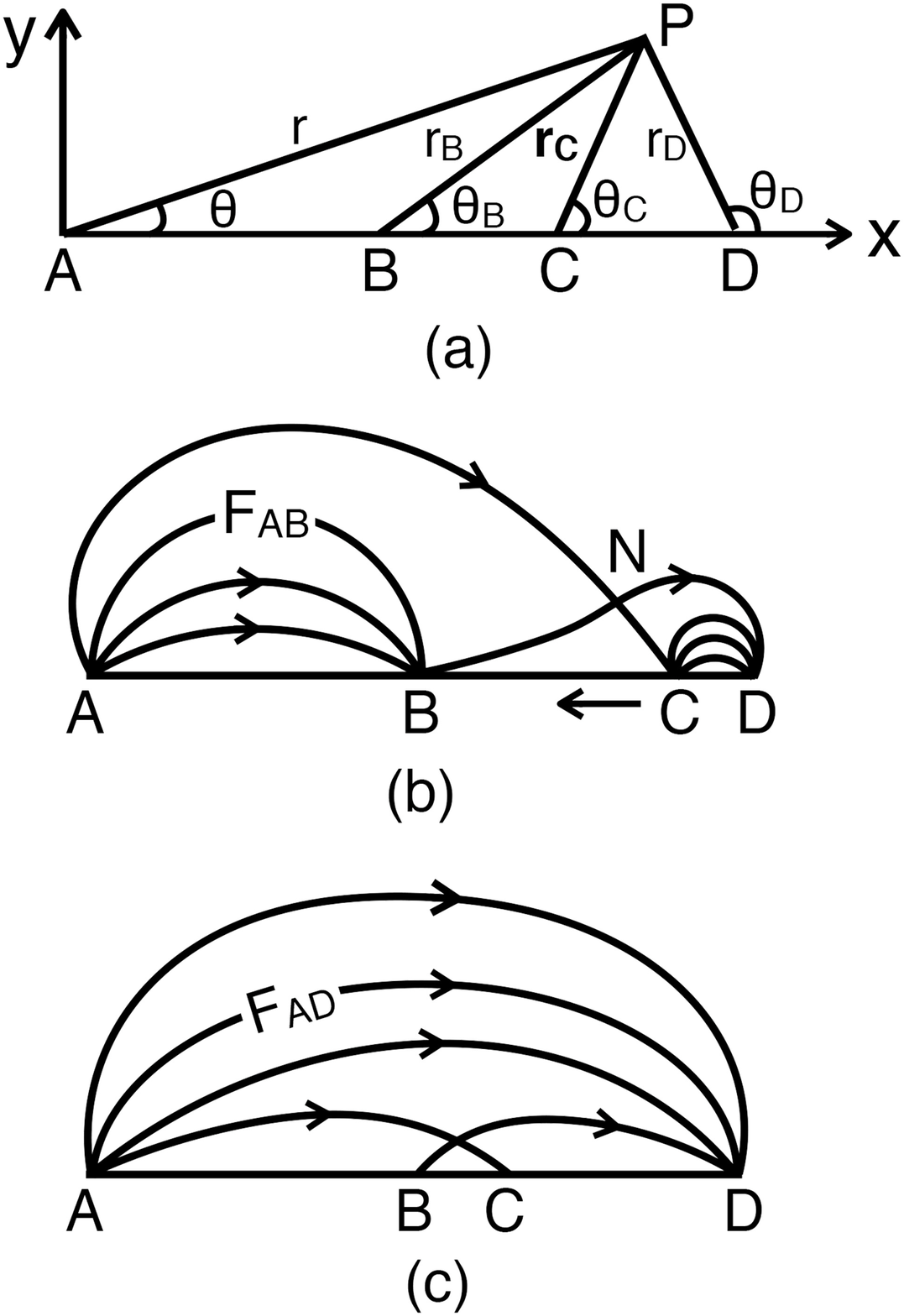}
\caption{(a) The notation and a sketch of magnetic field lines due to two positive (A and C) and two negative (B and D) sources located on the $x$-axis when (b) C is close to D and (c) C is close to B, where the fluxes joining A to B  and C to D are $F_{AB}$ and $F_{CD}$, respectively, }
\label{fig6}}
\end{figure}

\begin{figure}[h]
{\centering
 \includegraphics[width=16cm]{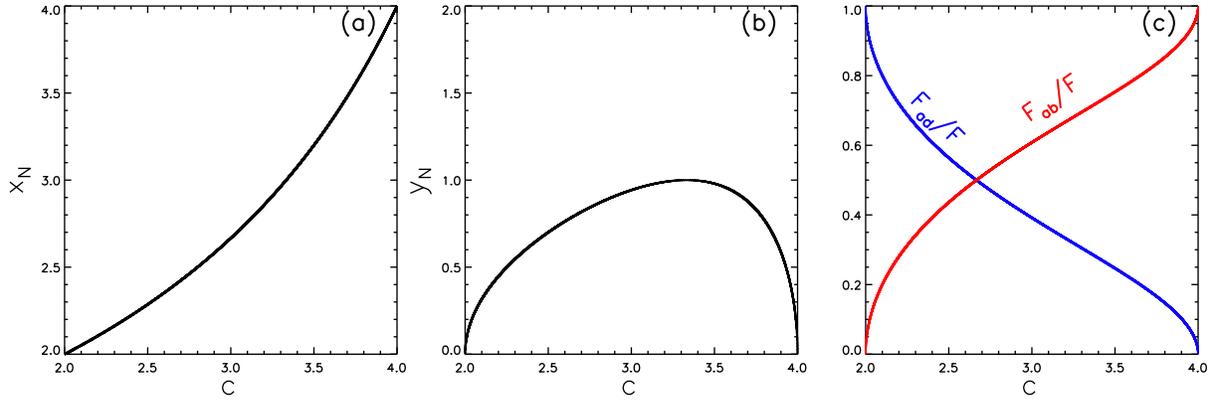}
\caption{The way the (a) $x$-coordinate and (b) $y$-coordinate of the null point and the magnetic fluxes ($F_{AB},F_{AD}$) joining A to B and A to D vary with the parameter $c$ as the point C moves from D to B.}
\label{fig7}}
\end{figure}

\begin{figure}[h]
{\centering
 \includegraphics[width=12cm]{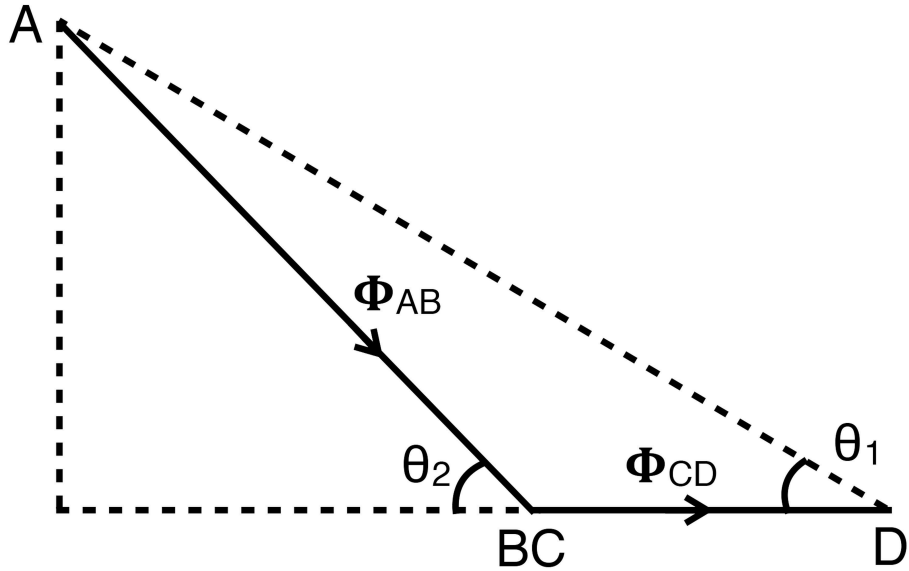}
\caption{The notation for the reconnection of two tubes AB and CD (of twist $\Phi_{AB}$ and $\Phi_{CD}$, respectively) to form a single tube AD.}
\label{fig8}}
\end{figure}

\begin{figure}
\epsscale{.80}
\plotone{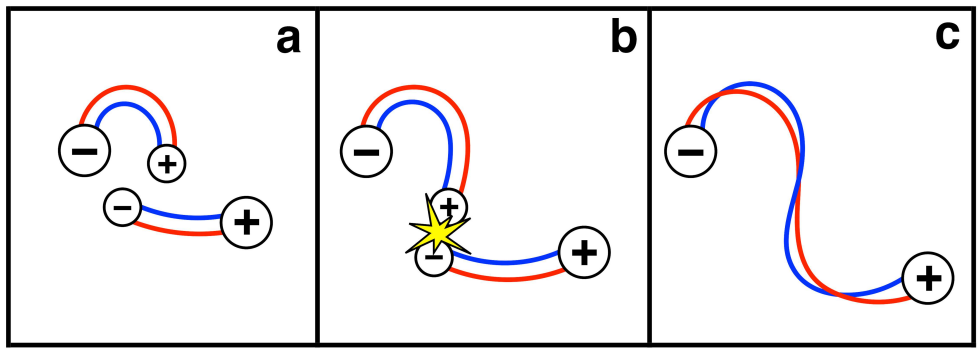}
\caption{Schematic picture showing the formation of the inverse S-shaped active-region filament. \label{fig1}}
\end{figure}
%\begin{figure}
%\centering
%\includegraphics[angle=0,scale=0.5]{fig2-1.eps}\\
%\includegraphics[angle=0,scale=0.49]{fig2-2.eps}\\
%\includegraphics[angle=0,scale=0.5]{cross1.eps}\\
%\includegraphics[angle=0,scale=0.5]{cross2.eps}\\
%\caption{Upper panel: The 3-D evolution of the penumbra and umbra of the sunspots in the active region NOAA 11884.  Lower panel: The projection of the upper panel of the Fig. 2 seen from the top to the photospheric surface. The X and Y axises of Fig. 2 show the field of view in pixel, which is the same as field of view of Fig. 1. The Z axis of the upper panel of Fig. 2 denote the number of the continuum intensity images from 0 to 480 frames. The corresponding time is from 00:00:00 on October 31 to 23:48:00 on November 3, 2013. The color bar shows the intensity of the bottom image in Fig. 2. The red, the blue and the white curves denote the evolution of the sunspot S2, the evolution of the sunspots P2 and P1, respectively. The green and the purple arrows indicate the penumbra and umbra of the sunspots in the figure. }
%\end{figure}%\begin{figure}

\end{document}